\documentclass[12pt,a4paper]{article}
\usepackage{epsfig}
\usepackage{ajax}

\newcounter{appex}
\newcommand{\appex}[1]{\refstepcounter{appex}%
\setcounter{equation}{0}\setcounter{subsection}{0}%
\section*{{\it Appendix} \Ajax{appex}\quad #1}%
\addcontentsline{toc}{section}{{\it Приложение} \Ajax{appex}\quad
#1}}
\newenvironment{append}%
{  \setcounter{appex}{0}%
}%
{ 
}

\begin{document}

\begin{center}
{\bf LONG MEMORY IN STOCK TRADING }
\end{center}

\bigskip

\begin{center}
{\bf Andrei Leonidov}
\end{center}

\begin{center}

\medskip

{\it (a) Theoretical Physics Department, P.N. Lebedev Physics Institute,\\
119991 Leninsky pr. 53, Moscow, Russia}\footnote{Research
supported by RFBR grant 02-02-16779 and Scientific School Supprort
grant 1936.2003.02 }

\medskip

{\it (b) Netcominvest Financial Investment Company,\\ 109017 Profsoyznaya 3, Moscow, Russia}
\footnote{Address before May 1 2003: Quantitative Analysis Group, EFOT  GmbH, Feuerbachstrasse 26-32,
60325 Frankfurt, Germany}\\

\medskip

{\it (c) Institute of Theoretical and Experimental Physics\\
117259 B. Cheremushkinskaya 25, Moscow, Russia}
\end{center}

\bigskip

\begin{abstract}
Using a relationship between the moments of the probability distribution of times between the two
consecutive trades (intertrade time distribution) and the moments of the distribution
of a daily number of trades, we show that the underlying point process is essentially non-Markovian.
A detailed analysis of all trades in the EESR stock on the Moscow International Currency Exchange in
the period January 2003 - September 2003, including correlation between intertrade time intervals is presented.
A power-law decay of the correlation function provides an additional evidence of the long-memory nature of the
series of times of trades. A data set including all trades in Siemens, Commerzbank and Karstadt stocks traded
on the Xetra electronic stock exchange of Deutsche Boerse in October 2002 is also considered.
\end{abstract}

\newpage

The new field of econophysics, in particular part of these activities related to the description of financial markets,
has drawn considerable attention in recent years \cite{MS00,BP01,S03}.  One of the central issues under discussion
is understanding fundamental origins of the stock price formation process. The progress in this direction is very rapid,
see, e.g., \cite{PGAGS00,GGPS03,BGPW03,LF03,FGLMS03,WR03,WR04} and references therein. The ultimate goal of the quantitative
approach to the description of stock price dynamics is constructing a microscopic theory, in which this dynamics originates
from the agents' activity in the financial market generating patterns of demand - supply interaction, resulting in price
formation. At a more phenomenological level the description of stock price evolution focuses on the dynamics of accomplished
trades.

Among the most important characteristics of financial markets is their activity. On the trade-by-trade
basis one can think of two possible ways of characterizing it and thus setting the clock measuring
the operational time. The first possibility is to follow the volume traded \cite{C73}, the second is to analyze
the time-dependent frequency of trading operations. Below we shall follow the latter route.
The importance of effects related to the varying trading frequency is well recognized. In particular, in
\cite{D02} it was suggested, that they significantly influence the expected return during intensive speculative
trading. In \cite{PGAGS00} it was also argued that the empirically observed long-range correlations in trading frequency
\cite{BLM00,PGAGS00} directly induce the observed long-range correlations of volatility. Let us also
mention a related topic of great practical interest - a study of seasonality effects
(e.g., intraday and intraweek activity patterns) and of the corresponding tuning of trading strategies, see e.g.
\cite{HFF01}.

Below we shall study the trade dynamics at the smallest level of resolution, focusing ourselves on the
properties of a (stochastic) process generating the times of trades. Given a description of such {\it directing} point
process $T(t)$ generating the times of trades $t_1 \leq t_2  .....$, the temporal evolution of quantities of interest
such as price or volume is that of a {\it subordinated} stochastic process \cite{F66} $P(T(t))$. The form the
corresponding evolution equation for the probability distributions of, e.g., price $\Omega (P,t)$ and
its characteristic features are thus decisively dependent on the properties of the directing point process $T(t)$.
Generically one can distinguish the following possibilities:

\begin{itemize}

\item{{\bf Type 1}. The time at which the $n$-th point (trade) takes place is completely independent of the times at
which the previous $n-1$ points were generated. In this case one deals with the Poisson distribution of the number of
trades in any fixed time interval. This point process is fully characterized by an {\it exponential} probability
distribution for the intertrade time $\tau=t_n-t_{n-1}$, $\psi(\tau)=\tau_0^{-1} {\rm exp} (-\tau/\tau_0)$. The evolution
of the probability density $\Omega (P,t)$ is described by a differential (in time) equation.}

\item{{\bf Type 2}. The time of the $n$-th trade is correlated with that of the previous, $n-1$-th, trade, so that the the
point process has a unit memory depth (and is thus still Markovian). The point process $T(t)$ is still fully characterized
by some {\it non-exponential} intertrade time probability distribution $\psi(\tau)$, but the evolution equation for
$\Omega (P,t)$  is no longer differential, but an integral one, of continuous-time random walk type \cite{MS84}
\footnote{The application of CTRW to the analysis of financial data was discussed in a number of papers
\cite{SGM00,MRGS00,RSM02,MMG02,MMPW03,SGMMR03,RR03}}. The correlation between the number of trades in
non-overlapping time windows can in principle be computed \cite{D50}, but the resulting expressions
are quite involved.}

\item{{\bf Type 3}. The time of the n-th trade depends on $r>1$ times of the previous trades
$t_{n-1}, t_{n-2}, ..., t_{n-r}$. In this case the point process is a long-range memory non-Markovian one with a memory
depth equal to $r$, and the corresponding evolution equation for $\Omega (P,t)$ is a complicated integral one with a kernel
depending on $r-1$ arguments. If all temporal scales are effectively involved, $r$ is infinite.}

\end{itemize}

\begin{figure}[h]
 \begin{center}
 \epsfig{file=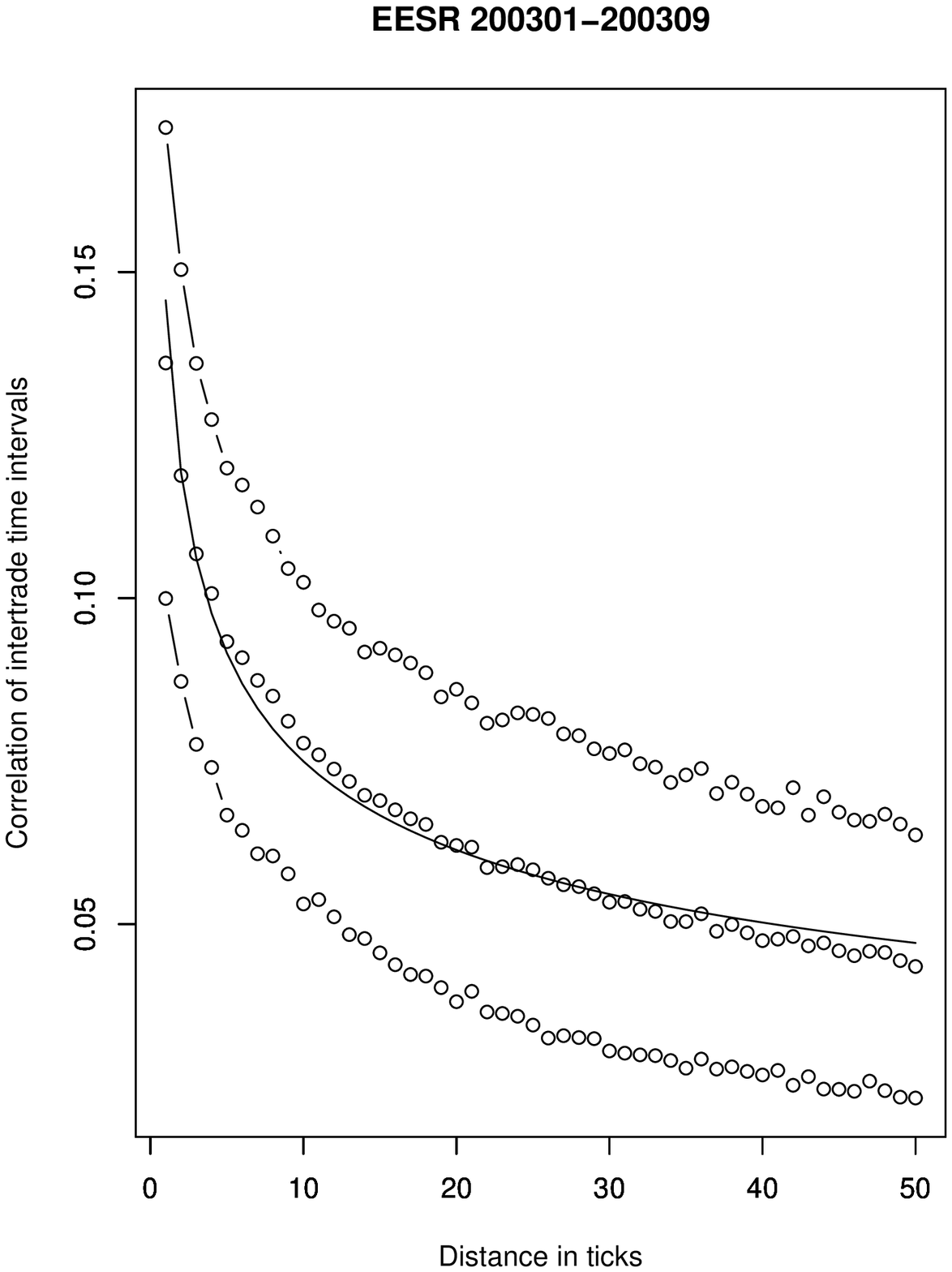,height=10cm,width=14cm}
 \end{center}
 \caption{Normalized correlation of intertrade time intervals in tick time. Solid line shows a powerlike
 fit of the form $0.15/d^{0.29}$. Upper and lower lines show the standard deviation of the measured correlation
 characterizing its variation on the day-to-day basis.}
 \label{tcor}
\end{figure}

The main question is, of course, whether the times of trades correspond to a long-range memory process with past events
strongly influencing the new ones. Let us first consider a standard, although not very reliable, measure of the
amount of memory present in the series - the (normalized) correlation function of the intertrade time intervals
\begin{equation}\label{ecor}
     C_{\tau}(d) \, = \, { \overline{\langle \tau_k \tau_{k+d} \rangle_k} -\langle \tau \rangle ^2 \over
     \langle \tau^2 \rangle}
\end{equation}
The correlation is computed by first averaging over all intervals separated by a fixed number of intervals $d$ for a
given day and then averaging over all days. The correlation (\ref{ecor}) for all trades, from January 2003 till September
2003, in the EESR stock traded on the MICEX stock exchange is shown, together with the errorbars
characterizing its variation on the day-to-day basis, in Fig.~\ref{tcor}.
The correlation function $C_{\tau}(d)$ shown in Fig.~\ref{tcor} is clearly powerlike, $C_{\tau}(d) \simeq 0.15/d^{0.29}$,
thus demonstrating a behavior typical for long-range processes.

To get a somewhat more precise picture of the memory - related pattern of the set of times of trades, let us consider a
relation between the moments of the distribution $p_N (\Delta T)$ of a number of trades in some given time interval
$\Delta T$ and those of the intertrade time distribution $\psi (\tau)$ that holds for the unit memory depth
point process in the large -- $\Delta T$ limit \cite{D50}. More specifically, for the processes of Type 1 and Type 2
there exist, for a sufficiently large time interval $\Delta T$, a relation between the first two moments of the
distribution of the number of trades within this interval $\langle N_{\Delta T} \rangle $ and
$\langle N_{\Delta T}^2 \rangle $ and the first two moments of the intertrade time distribution
$\psi_1 \equiv \langle \tau \rangle $ and $\psi_2 \equiv \langle \tau^2 \rangle $ \cite{D50}:
\begin{equation}\label{ratio1}
 \langle N_{\Delta T}^2 \rangle  - \langle N_{\Delta T} \rangle ^2 = { \psi_2 - \psi_1^2 \over \psi_1^2 } \,
 {\Delta T \over \psi_1} \equiv
 { \psi_2 - \psi_1^2 \over \psi_1^2 } \, \langle N_{\Delta T} \rangle
\end{equation}
Corrections to the relation (\ref{ratio1}) are of order $1/\Delta T$. Equivalently,
\begin{equation}\label{ratio2}
 \rho_N \equiv {\langle N_{\Delta T}^2 \rangle - \langle N_{\Delta T} \rangle ^2 \over \langle N_{\Delta T} \rangle }
 = { \psi_2 - \psi_1^2 \over \psi_1^2 } \equiv \rho_{\tau}
\end{equation}
so that if the underlying point process $T(t)$ is indeed of Type 1 or Type 2, one should have $\rho_N/\rho_{\tau} = 1$.
Let us note, that in the simplest poissonian case $\rho_N = \rho_{\tau} = 1$. Let us also stress that provided the underlying
point process has unit memory length, the equation (\ref{ratio2}) holds for any intertrade time probability distribution
$\psi(\tau)$ having a finite second moment $\psi_2$. Derivation of Eq.~(\ref{ratio2}) is sketched in the Appendix.

Equation (\ref{ratio2}) provides a straightforward possibility of establishing the nature of the point process $T(t)$ by
computing the quantities in its left- and right-hand side from an observed distribution of a number of trades in some
fixed interval $\Delta T$ and a corresponding underlying distribution of the intertrade times. In particular in the case
of the ensemble of independent points (Poisson case) $\rho_N = \rho_{\tau} = 1$, in the case of the CTRW - type process
having unit memory depth a more general relation of Eq.~(\ref{ratio2}) holds and, finally, for  the long-range memory
process the relation in Eq.~(\ref{ratio2}) should break down.

In the data set used for our analysis the interval $\Delta T$ corresponds to one trading day (10.30 - 18.45 on the
MICEX exchange and 9.00 -- 20.00 on the Xetra electronic exchange of Deutsche Boerse  in Frankfurt), and the considered
ensemble of trades consists of all trades in EESR in January - September 2003 (MICEX) and  Siemens, Commerzbank
and Karstadt (liquid, medium liquid and relatively illiquid stock) in October 2002 on Xetra. The results are summarized
in Table 1:

\bigskip

\begin{center}
{\bf Table 1}

\medskip

\begin{tabular}{|c|c|c|c|c|c|}
\hline
 Und & $\langle N_{\Delta T} \rangle$ & $\langle \tau \rangle$ (seq)& $\rho_{\tau}$ & $\rho_N$ & $\rho_N/\rho_{\tau}$ \\
\hline
 EESR & 6823 & 4.5 & 10.3 & 492.8 & 109.4 \\
\hline
 SIE & 4364 & 9.1 & 3.8 & 124.8 & 32.8 \\
\hline
 CBK & 1856 & 21.3 & 3.1 & 274.4 & 88.5 \\
\hline
 KAR & 373 & 104.1 & 4.5 & 39.7 & 8.8 \\
\hline
\end{tabular}
\end{center}

From the results presented in Table 1 it is clear, that poissonian Type 1 and CTRW Type 2 processes are excluded as
candidate point processes generating the times of the trades. This leaves us with the only remaining possibility of
Type 3 non-Markovian long-range memory process. This conclusion is in obvious agreement with the long-range correlation
between the intertrade time intervals for EESR shown in Fig.~(\ref{tcor}).

To provide some statistical backing to this conclusion, we have generated the times of trades using the empirically
observed distribution over intertrade times as the EESR data. The thus generated set of points is characterized by the
observed distribution, but is otherwise uncorrelated \footnote{The distributions over intertrade time intervals were
studied in a number of papers \cite{RSM02,SGMMR03,RR03}}. The considered number of trade intervals $\Delta T$ was
equal to the number of trading days in the EESR data. This simulation effectively reconstructs a unit memory depth CTRW
process. The results of this simulation are shown in Table 2:

\begin{center}
{\bf Table 2}

\medskip

\begin{tabular}{|c|c|c|c|c|c|}
\hline
 Und & $\langle N_{\Delta T} \rangle$ & $\langle \tau \rangle$ (seq)& $\rho_{\tau}$ & $\rho_N$ & $\rho_N/\rho_{\tau}$ \\
\hline
 EESR (data)& 6823 & 4.5 & 10.3 & 492.8 & 109.4 \\
\hline
 EESR (sim.) & 6816 & 4.5 & 12.4 & 8.2 & 0.66 \\
\hline
\end{tabular}
\end{center}
We see, that although due to statistical limitations (finite size corrections, etc.) the simulation does not reproduce
the theoretical value of $\rho_N/\rho_{\tau}=1$, it is sufficiently close to it, while at the same time missing the
experimental ratio by two orders of magnitude.

\begin{center}
{\bf Conclusion}
\end{center}

We conclude that the point process generating the times of trades is of long-range memory non-Markovian
nature. This excludes random walk and continuous-time random walk (CTRW) processes as models describing
high frequency stock price dynamics in real time.

\begin{center}
{\bf Acknowledgements}
\end{center}

I am grateful to Alain Guillot, Stefan Iglesias, Larry McLerran and Axel Vischer for reading the manuscript and useful
comments. I would also like to thank the referee for constructive and helpful remarks.

\begin{append}
\appex{Derivation of Eq.~(3)}%

In this Appendix we sketch, following \cite{D50}, the derivation of Eq.~(\ref{ratio2}). The probability of
having $N$ trades in the interval $[0,T]$ is conviniently written as a convolution of a probability of having
$N$ trades in the interval $[0,t]$ (the $N$-th trade falling into the infinitesimal vicinity of $t$) $w_N (t)$
and a probability of having no trades in the remaining interval $[t,T]$ $w_0 (T-t)$:
\begin{equation}\label{A1}
 p_N (T) \, = \, \int_0^T dt\, w_0 (T-t)\, w_N (t)
\end{equation}
so that the Laplace transform of $p_N (T)$ reads
\begin{equation}\label{A2}
 {\tilde p}_N (\lambda) \, = \, \int dT\, {\rm e}^{-\lambda T} p_N(T) \, = \,
 {\tilde w}_0(\lambda)\,{\tilde w}_N(\lambda) \, = \,
 { 1 - {\tilde \psi} (\lambda) \over \lambda}\, {\tilde \psi} (\lambda)^{N-1},
\end{equation}
where we took into account that in the considered unit memory depth case all probabilities in question are fully
determined by the intertrade (waiting) time distribution $\psi(\tau)$, so that
\footnote{Here we assume that there is always a first trade at the beginning of the time interval in question}
${\tilde w}_N(\lambda)= {\tilde w}(\lambda)^{N-1}$ and, for $w_0$ we have, recalling  that
$w_0(\tau)\,=\,\int_{\tau}^\infty d \tau'\, \psi(\tau')$,
${\tilde w}_0 (\lambda) \, = \, (1 - {\tilde \psi} (\lambda))/\lambda$.

To compute the moments of trade's  multiplicity distribution it is
convenient to introduce a generating function
\begin{equation}\label{A3}
G(\theta | T) \, = \, \sum_{k=1}^{\infty} {\rm e}^{k \theta} p_k (T)
\, = \, 1 + \sum_{k=1}^{\infty} {\theta^k \over k !} \langle N^k \rangle
\end{equation}
so that the moments are given by its derivatives at the origin
\begin{equation}\label{A4}
 \langle N^k \rangle_T \, = \, \left ( {\partial^k G \over  \partial \theta^k }
\right )_{\theta=0}.
\end{equation}

Combining Eqs~(\ref{A2},\ref{A3}) one gets the following expression for
the generating function:
\begin{equation}\label{A5}
G(\theta | T) \, = \, {1 \over 2\pi i} \int_{c-i\infty}^{c+i\infty} d \lambda \,
{\rm e}^{\lambda T}\, {1 \over \lambda} \, {1-\psi(\lambda) \over {\rm e}^{-\theta}+\psi(\lambda)}
\end{equation}
Equation (\ref{ratio2}) follows now by substituting the expansion of $\psi(\lambda)$ in
the moments of $\psi(\tau)$
\begin{equation}
\psi(\lambda) \, = \, \int_0^\infty d \tau \, {\rm e}^{-\lambda \tau}\, \psi(\tau)
\, = \, 1 - \psi_1\,\lambda + {1 \over 2}\, \psi_2 \, \lambda^2 + ...
\end{equation}
into Eq.~(\ref{A5}) and computing, through Eq.~(\ref{A4}), the leading contributions to
$\langle N_T \rangle$ and $\langle N_T^2 \rangle$ in the large $T$ limit
\begin{eqnarray}
 \langle N \rangle_T & \simeq & {T \over \psi_1} + {1 \over 2}\,{\psi_2 \over \psi_1^2} \\
 \langle N^2 \rangle_T & \simeq & {T^2 \over \psi_1^2} + 2T {\psi_2 \over \psi_1^3} - {T \over \psi_1}
\end{eqnarray}

\end{append}

\end{document}